%% file: wicsa2001.tex
\documentclass[10pt,twocolumn]{article} 
\usepackage{generic}
\usepackage{times,mathptm}
\usepackage{graphicx}
\usepackage{latex8}
\usepackage{fancyheadings}
\usepackage{kun-institute}
\usepackage{url}

\begin{document}

\title{Developing an Architecture Method Library}

\author{
     R.D.T. Janssen$^1$, 
     H.A. Proper$^{1,2}$,
     H. Bosma$^1$, D. Verhoef$^1$,
     S.J.B.A. Hoppenbrouwers$^2$
}
\email{\rm
     \{R.Janssen, E.Proper, H.Bosma, D.Verhoef\}@ordina.institute.nl,
     stijnh@cs.kun.nl
}
\affiliation{\it
     $^1$Ordina Institute, Groningenweg 6, 2803 PV Gouda, The Netherlands \\
	     \it $^2$University of Nijmegen, IRIS,
             Toernooiveld 1, 6525 ED Nijmegen, The Netherlands }

\maketitle
\thispagestyle{fancy}
\setlength{\headrulewidth}{0pt}
\cfoot{\textsf{\small
  Submitted to: WICSA2001, Working IEEE/IFIP Conference
  on Software Architecture, Amsterdam, August 28--31, 2001.}}

\input{pub}

\begin{abstract}
  Today, there are millions of professionals worldwide acting as a designer,
  architect or engineer in the design, realization, and implementation of
  information systems. At this moment there is no 
  well established and clearly identified
  body of knowledge that defines their profession in a ``standard'' way.

  In this article, we present the idea of developing an architecture
  method library. Such a library could play a pivotal role to 
  further professionalize the field.
  The library contains project experiences, reference
  architectures, literature, proven methods, tools, etc.

  Access mechanisms allow the professional 
  to use this body of knowledge. By giving it an open nature, it can
  be filled by professionals from different fields. Feedback mechanisms are 
  possible to improve the contents of the library, for example by 
  giving feedback on the method components in the library. 
\end{abstract}

\Section{Introduction}
The design, realization, and implementation of
information systems provides employment to millions of professionals worldwide. 
When attempting to clearly identify their profession, one discovers
that there does not exist a well established and clearly identified
body of knowledge.

The Software Engineering Body of Knowledge (SWEBOK)
project~\cite{WWW:SWEBOK:SoftwareEngineering} started from a similar
observation for the field of software engineering. 
The following motivation of the SWEBOK project is provided: 
\begin{quote}
  In other engineering disciplines, the accreditation 
  of university curricula and the
  licensing and certification of practicing professionals are 
  taken very seriously.
  These activities are seen as critical to the constant upgrading 
  of professionals and,
  hence, the improvement of the level of professional practice.
  Recognizing a core body of knowledge is pivotal 
  to the development and accreditation
  of university curricula and the licensing and certification of professionals.
\end{quote}
This inspired us to develop an architecture method library,
containing project experiences, reference
architectures, literature, proven methods, tools, etc.

Using various kinds of access mechanisms the professional 
is able to use this library. By giving it an open nature, it can
be filled by professionals from different fields and with different
backgrounds. Feedback mechanisms can 
be used to improve the contents of the library by e.g. giving feedback 
on the method components present based on the actual use of the library.

This article describes the design of such an architecture method
library, what it contains, the way it is filled and how it can be used. 
Some future directions are given. In section~\ref{sec:example}
an extended example is given to illustrate the results
from the other sections.

\Section{Restrictions on the scope of the library}
Given the unclear boundaries of the information systems engineering
field (what is part of this domain and what is not?),
we use the following definition of information system: 
\begin{quote} 
   An \emph{information system} is a sub-system of an organizational system, 
   comprising the communication- and information-oriented 
   aspects of the organization system~\cite{Report:90:Lindgreen:FRISCO}.
\end{quote}

In this article, the term information systems engineering
is used to refer to all activities involved in the
design, realization and implementation of information systems
(including non-IT or non-automated aspects). This
essentially ranges from the high level design of a portfolio of information
systems (an information architecture) to the design and realization of a
specific information system.

We use an organic approach to the development of the architecture method
library. This essentially means that no strong restrictions are set on the
scope of the library beforehand.
It will emerge from the content as more is added.
This fits with the fact that most of the knowledge available on
information systems engineering is in case-based form,
tied to experiences learned from specific cases from industrial practice.

For this moment, the demarcation lies in the group
of professionals responsible for providing the content: those people
who consider themselves to be architects or information systems engineers. 

We realize that we cannot aim to develop a fully integrated and completely
consistent architecture method library. 
Reasons for this are (among others) the huge amount of 
literature in the information
systems field, the `vagueness' of the field (and,
subsequently, the
various backgrounds of the professionals involved), and the case based nature.
Therefore, we aim to compile and structure a library
filled with a collection of significant, loosely coupled `knowledge
items' or `method components',
considered useful to project members of an information systems
engineering project. These method components consist of focused
denotations of knowledge pertaining to information systems engineering.

\Section{Related work}
Related work can be found in e.g.~\cite{Book:97:IS97:Curriculum}
and~\cite{Report:90:Lindgreen:FRISCO}.
In~\cite{Book:97:IS97:Curriculum}, a model curriculum is described for
undergraduate degree programs in information systems and it also
contains a definition of a body of knowledge for information systems
engineering. The Framework for Information Systems Concepts (FRISCO) as
reported in~\cite{Report:90:Lindgreen:FRISCO} aims to give the field of
information systems a conceptual underpinning by introducing a unified
framework of concepts. What both of these approaches lack is the
practical side in terms of concrete work practices, techniques and tools
to be used, etc.

Another field that is of relevance is the field of method
engineering~\cite{Book:96:Brinkkemper:MethodEngineering,PhdThesis:97:Harmsen:MethodEngineering}.
The approach taken there tends to dissect methods to
the level of distinct concepts and their relationships. For the moment,
it involves too much detail to be applied to the gathering
and structuring of a library of knowledge items.
However, the theories provided by method engineering are useful in
dissecting method knowledge into more elementary items
and will also be useful for developing a 
consistent body of method knowledge once a significant
collection of knowledge items has been gathered into our architecture
method library.

\Section{Contents of the library}
The architecture method library should contain 
information about various architecture and information systems 
subjects, such as:
\begin{itemize}
  \item project strategies,
  \item case studies (both successes and failures),
  \item descriptions of methods, techniques and tools,
  \item reference models and frameworks,
  \item heuristics on the use of project strategies, methods,
    techniques, etc., 
  \item knowledge to aid in planning and execution of projects,
  \item heuristics and guidelines with regard to design decisions, the use
        of reference models, architecture principles, etc.
\end{itemize}

The intended audience of the library includes managers, architects,
designers, engineers, etc.  In the next sections it is shown how we
intend to dissect this knowledge into focused `knowledge items' or `method
components' as we will call them below.

First, method components and a characterization of them are introduced.
This is followed by situational factors and heuristics.
Finally, the complete structure of the library is given.

\SubSection{Method components}
\label{sec:methodcomp}
In literature on method engineering, frameworks are found that
basically provide an anatomy of a method. See for 
example~\cite{Book:96:Fokkinga:LAD,Article:90:Wijers:Modelling}. This
has led us to define our framework for
building an architecture method library on five ``standard'' questions,
which can be identified with the following method components:
\begin{itemize}
  \item
    \emph{Product.} 
    Based on the question ``what has to be delivered?''
  \item
    \emph{Activity.}
    Based on the question
    ``what has to be done to obtain the deliverable?''
  \item
    \emph{Actor.} 
    Based on the question
    ``who will be doing it and what capabilities do they need?''
  \item
    \emph{Tool.} 
    Based on the question
    ``what tools can be used?''
  \item
    \emph{Principle.} 
    Based on the question
    ``what architecture principles have to be used?''
\end{itemize}

\SubSection{Characterization of method components}
\label{sec:props}
Each method component has to be characterized so that a potential
user is able to judge whether that component may be of use to him. For
example, for a modeling technique, it is important to know for what
types of modeling it may be used for: e.g. for modeling of business
processes, for modeling of organizational structures, for communication, etc.

For this purpose, we define \emph{properties} with \emph{property values}.
During the development of the
library it will be discovered which of them are useful and which not. Examples
of such properties are:
\begin{itemize}
  \item \emph{System scope.}
	What is the scope of the method component in terms of
	the system(s) considered?
        Some examples are: portfolio of systems, family of systems,
	specific system, and specific use-case.
  \item \emph{Systemic focus.}
        What system is the method component focused at?
	Some options are:
	organizational system, information system, computerized
	information system, and infrastructural system.
  \item \emph{Systemic aspect.}
	Zachman has identified a number of
	different aspects of a system~\cite{Article:87:Zachman:ISA}, 
	based on the interrogatives of
	the English language.
	This leads to aspects such as: what, how, when, why, and where.
  \item \emph{Level of abstraction.}
        What level of abstraction, with respect to the systems under
        consideration, is the method component focused at? Examples
        (derived from~\cite{Report:82:ISO:Concepts}) are: conceptual level,
        logical level, and physical level.
  \item \emph{Communication style.}
	What is the communication style used in describing the
	method component? Some examples are: reference material,
	study material, and teaching material.
  \item \emph{Natural language used.}
	What is the language used to express the method component?
	Examples are: Dutch, English, American English, German,
	and French.
  \item \emph{Development cycle.}
        Where in the development cycle can the method
	component be positioned? 
	For example, activities can belong to the following aspects:
        delivery approach, construction approach, cognitive approach, 
        social approach, development approach, quality control, and
        configuration control~\cite{Book:99:ISPL:ISPL-Introduction}.
\end{itemize}

\begin{figure}
\begin{center}
\includegraphics[width=7cm]{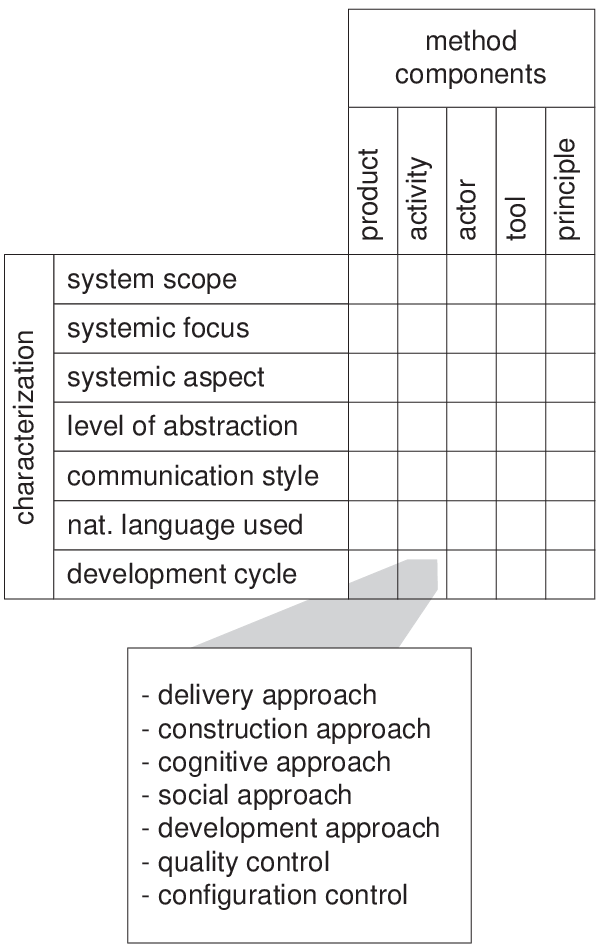}
\end{center}
\caption{The method component characterization can be seen as 
being orthogonal to
the method components. The aspects of `activities' of the
characterization `development cycle' are shown as example.} 
\label{fig:characterization}
\end{figure}

The characterization can be seen as being 
orthogonal to the method components. An
illustration is given in figure~\ref{fig:characterization}.

\SubSection{Situational factors}
\label{sec:sitfact}
For each method component in the library, there may exist various
situational factors which determine its applicability. 
In this context, a situational factor is defined as:
\begin{quote}
  A \emph{situational factor} is a property of the 
  problem situation that can be used to determine the
  most appropriate problem-solving strategy. This includes those
  properties that can have an impact on the different types of
  uncertain events which may occur and their adverse 
  consequences~\cite{ISPL-dictionary99}.
\end{quote}

A situational factor is important since
it determines how a method component can be used and how it can be
adapted to a
specific situation. And, vice versa, given a specific situation, the
situational factor indicates which method components have been proven to work
in that specific situation. 

For the library, this means that it is necessary for the user
to be able to select
method components based on the situation at hand. Therefore, there
has to be a way to specify the situational factors. This is done by
giving them a name and a value, e.g. 
\begin{center}
\emph{name} = CompanyType\\
\emph{value} = \{Financial company, Public company, Industrial company, ...\}
\end{center}

\SubSection{Heuristics}
\label{sec:heurs}
The situational factors are linked to the method components by
using heuristics. A heuristic is a rule-of-thumb specifying which
method component can be used in which situation. This is
represented using an IF...THEN... rule with a logical expression
about the project situation at hand in the IF part and the method component
to be used in that situation in the THEN part. For example:
\begin{center}
IF the complexity of the data to model is high \\
THEN it can be recommended to use a modeling technique based on natural
language
\end{center}
Another example based on the relation between two method components is:
\begin{center}
IF a participatory approach is used in the architects team \\
THEN it can be recommended to have a socially skillful team
\end{center}

\begin{figure}
\begin{center}
\includegraphics[width=8.25cm]{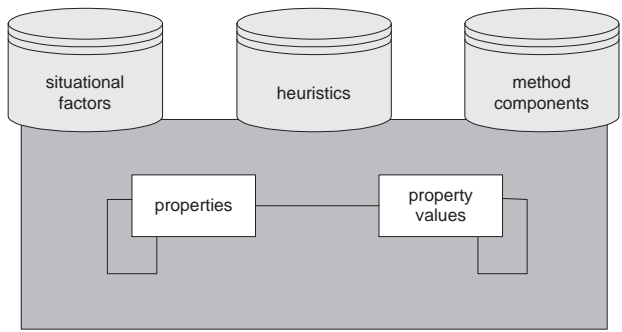}
\end{center}
\caption{The structure of the library.}
\label{fig:libstructure}
\end{figure}

\SubSection{Structure of the library}
The complete structure of the library is given in 
figure~\ref{fig:libstructure}. 
It is a collection of
situational factors, heuristics, activities, tools, etc. which are
coupled by relations. The heuristics specify which relations can be
applied when. The properties define the method components and 
are being used in the heuristics. 
In section~\ref{sec:example} (extended example), examples
of method components, attributes, and the relations between them 
can be found.

\Section{Using the library}
\label{sec:uselib}
Project members can use the architecture method
library to find the answers to questions such as: 
\begin{itemize}
\item 
  Which activities, techniques and tools are most appropriate given a
  specific project situation and task? 
\item
  What reference models are relevant for my project?
\item 
  What reference models are relevant in a situation where the
  business has selected ``customer intimacy''
  as in~\cite{Book:97:Treacy:BusinessStrategy} as their strategic focus? 
\item
  What are relevant architecture principles?
\item
  Which reference models represent a front/mid/back office architecture?
\item
  Which guidelines are useful for designing a data model?
\item
  What methods can be used for designing an application architecture in
  a process intensive environment?
\end{itemize}

Just like a regular (physical) library, the architecture method library
is meant to \emph{assist} professionals. 
That means that the content in the library may
help users in deciding, for example, which activities are appropriate in 
a certain project situation. The library will then present several
possibilities, but only judgment of the professional will make this usable
in a project context. As always, the expert is responsible for decisions
made.

When the architecture method library is used, the general 
steps for answering questions like those above are the following:
\begin{enumerate}
\item
  Analysis of the situation at hand.
\item
  Selection of base method components for the question(s) at hand.
\item
  Selection of other useful method components. This step can be repeated
  as desired.
\item
  Use of all the method components selected helps to complete the
  analysis.
\end{enumerate}

\begin{figure}
\begin{center}
\includegraphics[width=8.25cm]{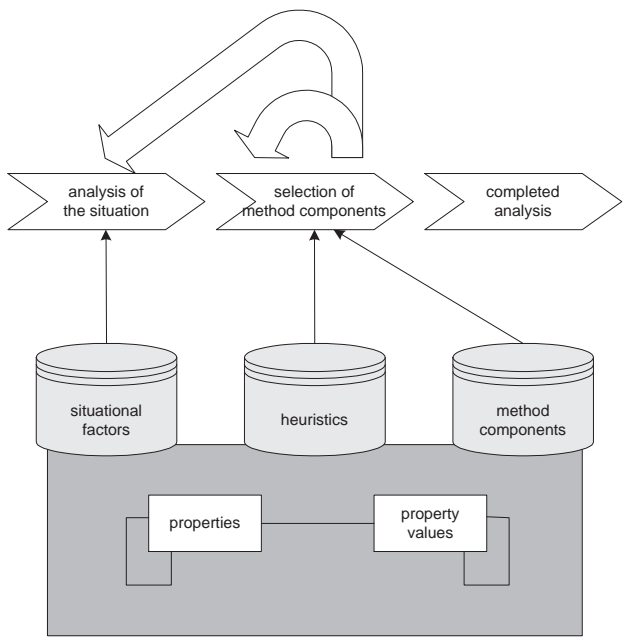}
\end{center}
\caption{An example use of the library.}
\label{fig:uselibrary}
\end{figure}

All this can be done by accessing the library. Adding these steps 
to the structure of the library as given in
figure~\ref{fig:libstructure}, the subsequent figure~\ref{fig:uselibrary}
results.

There are three different ways to access components of the library. The
first method is to formulate an exact \emph{query}. One formulates a question
based on logical expressions with values for various properties. This is
used as a query to the library. Examples of these kinds of queries are
given at the beginning of this section.

Often, it is not easy to formulate an exact query like this. In such
situations, browsing and selecting relevant method components
may be more appropriate. This is often called: \emph{query by
navigation}.

The third method of accessing the library is the use of a more-or-less
predefined \emph{decision tree}. Using such a tree as a guideline helps to
investigate the situation with which the user is confronted. The base
for this are situation factors and heuristics which are used to select
the applicable method components.

Of course all three access methods can be combined. After an initial
selection has been made, more can be added at will. This results in a
more refined analysis, as is indicated by the curved arrows
in figure~\ref{fig:uselibrary}. Method
components selected may (left open arrow) or may not (right open arrow)
lead to additional or revised analysis. 
In section~\ref{sec:example}, the different access mechanisms are
illustrated.

\Section{Filling the library}
\label{sec:filling}
Filling the library is an iterative process. That means that adding
new documents to the library is an ever continuing process. 

Examples of new, not yet analyzed documents
are project deliverables, project archives, textual
descriptions of methods, techniques and tools, case descriptions, etc.
From the perspective of the library, this material is not likely to be very
homogeneous with respect to the method components which are already
present. For example, a book
on some modeling method could discuss this technique in a way that is
highly intertwined with a specific tool that can be used to produce the
models which is already present in a method component in
the library. This relation can only
be added after (extensive) analysis of each source document.

After addition of these new documents, in the following step
this analysis is performed. 
From the source document, the method
components are extracted and described in a framework as described in
the previous section. During this step, 
the relations with other method components already in the library are 
determined. This is often a time consuming task.
The resulting material is homogeneous in nature.

In principle, anything can be added to the library. However, when
quality criteria are used for ``screening'' new raw material, the time
necessary for adding extracted method components can be reduced
drastically. Criteria can be:
\begin{itemize}
\item
  Is the new source document in some way structured (the structure may lead
  to easier identification of method components)?
\item
  Is the knowledge present in the new source document an addition to
  knowledge already present in the library (preventing double
  occurrences of the same method component)?
\item
  Does the material fit in the domain of the library (preventing the
  addition of irrelevant material)?
\item
  Is the knowledge expected to be
  interesting for more than one specific customer or
  situation (preventing the addition of material which is too specific to
  be used elsewhere)?
\end{itemize}

As can be seen, adding material to the library is a non-trivial task
which has to be done by an experienced architect. Guidelines are
important to prevent differences in interpretation of the raw material
and to prevent different phrasings. In the next section the importance
of this is illustrated. However, one should
realize that such differences cannot be prevented completely. 

Other, general
maintenance tasks for the library can be done by a qualified ``librarian''.

An additional step may be the rewriting of method components to some
form of ``standard terminology''. In this phase aggregation
can be executed, as well as normalization of terminology and concepts used.
However, this will take a substantial amount of time and the usefulness
of such an exercise is not clear at this moment.

\Section{Extended example}
\label{sec:example}
In this section an extended example is discussed.
We have taken~\cite{sanden97} as raw material.

\begin{figure*}
\begin{center}
\includegraphics[scale=0.9]{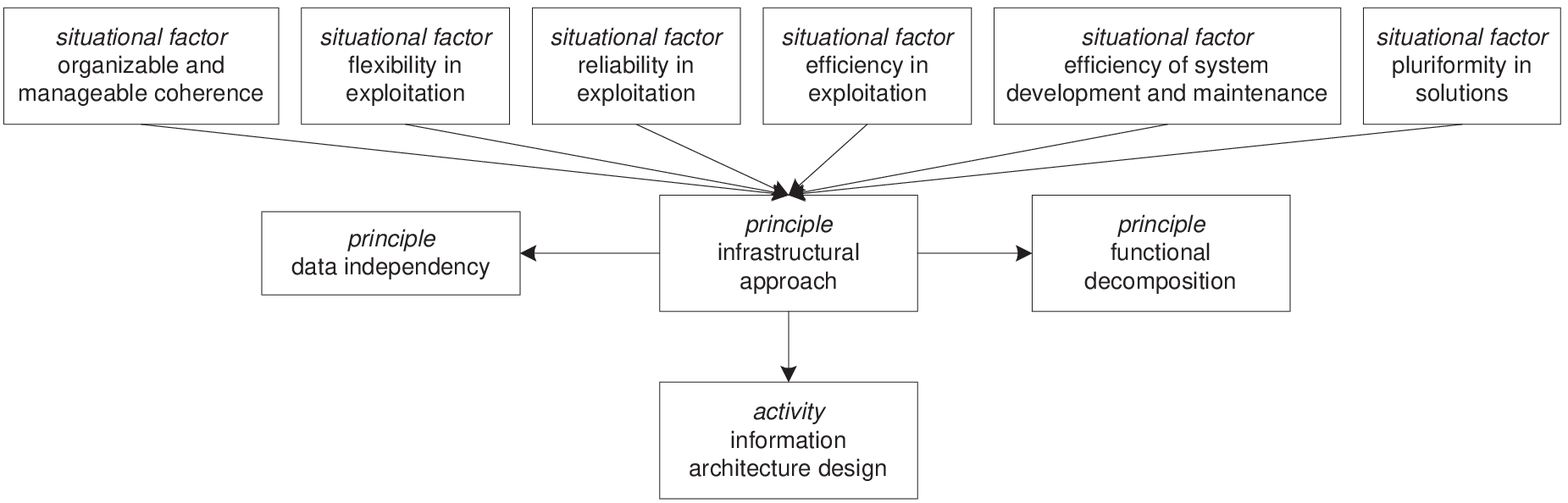}

\vspace*{6ex}

\includegraphics[scale=0.9]{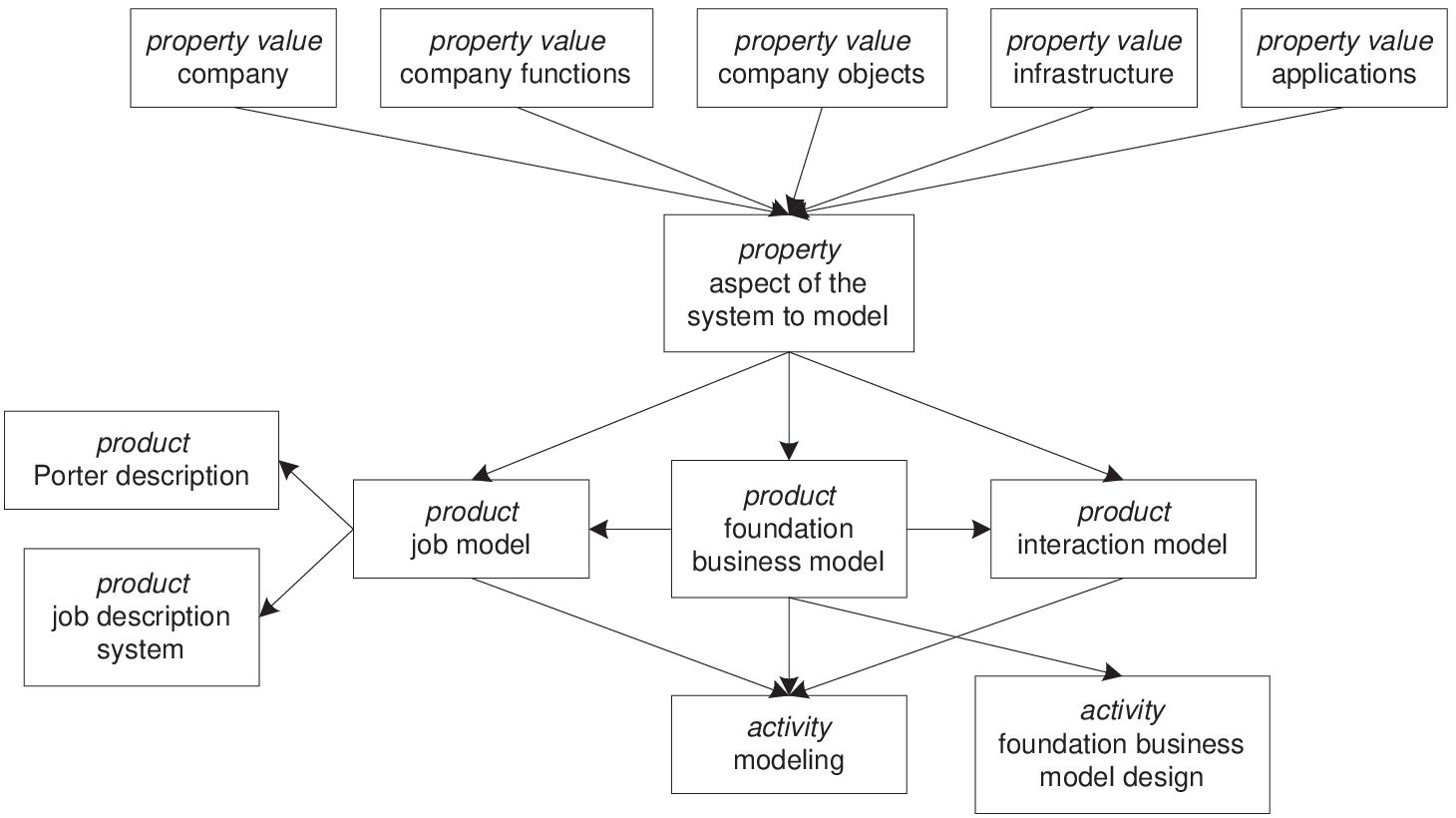}
\end{center}
\caption{Two examples with method components from~\cite{sanden97}. The
italicized line indicates 
the type of method component, and the other
lines the name, which in some cases can be found in the text. See
the text for further explanation.}
\label{fig:sandenexample}
\end{figure*}

\SubSection{Filling the library}
The analysis was started by checking the criteria from 
section~\ref{sec:filling}: we have found this book to be structured, the
knowledge was not in our architecture method library, the scope did fit
and since it is a general purpose architecture method book, it was certainly
interesting for more than one specific customer or situation.

First, we identified the various kinds of method components
(cf. sections~\ref{sec:methodcomp} to~\ref{sec:heurs}) 
and next the relations between them.

For each method component, a number of attributes was determined:
most importantly a (reference) name and 
a short description (which can be seen as a citation or short summary).
Then there are additional fields such as the reference
to the book, the page numbers, etc. The library
is flexible so more fields can be added as necessary.

Below, we will only show (for brevity) some of the method components 
identified from~\cite{sanden97}. Also, in
figure~\ref{fig:sandenexample} only a few of the
method components identified from this book are shown. 

\paragraph{Situational factors} ~ \\
Name: organizable and manageable coherence.\\
Description: coupling of systems determines the coherence of the
resulting information system. \\
Reference: \cite{sanden97}. \\
Page: 49.

Name: reliability in exploitation.\\
Description: relates to the capability of a
system to maintain its level of performance under stated conditions for
a stated period of time. \\
Reference: [ISO/IEC 9126 standard].

Name: efficiency in exploitation.\\
Description: relates to the relationship between
the level of performance of a system and the amount of resources used,
under stated conditions. \\
Reference: [ISO/IEC 9126 standard] (reference 
and pages for brevity omitted from now on).

\paragraph{Principle} ~ \\
Name: infrastructural approach.\\
Description: infrastructural rules guide the (partly) autonomous
processes. There are two layers: one with applications and one with
reusable software components. The infrastructure is common for both
layers.

Name: functional decomposition.\\
Description: functional decomposition allows applications to use common
centralized data and reusable software components.

\paragraph{Activity} ~ \\
Name: information architecture design. \\
Description: using the foundation business model, the infrastructural
components (function model and data maintenance system) and the business
organization (organization model) are determined. Using the organization
model and the system architecture the applications, data, and coherence
between them are developed.

\paragraph{Property} ~ \\
Name: aspect of the system to model. \\
Description: this is the perspective of the system to model.

\paragraph{Products} ~ \\
Name: foundation business model. \\
Description: global, independent model of the business. Contains
the purpose and a description of some sub models: job model, interaction
model and object model (last model not shown in the lower part of
figure~\ref{fig:sandenexample}).

\paragraph{Heuristics} ~ \\
In this example, heuristics are 
found in the upper part of figure~\ref{fig:sandenexample},
in the relations between the situational factors and
other method components (from the top most row
to the second row) (cf. section~\ref{sec:heurs}).

\paragraph{Relations between method components} ~ \\
The relations are shown as arrows in figure~\ref{fig:sandenexample}.

\paragraph{}The construction of other method components is similar.

\SubSection{Creating the network}
After identifying the method components and relations,
a network as shown in figure~\ref{fig:sandenexample}
can be constructed automatically.
Since for this example only one book was taken and not all method
components and relations have been drawn, it is only
a very small network. One can imagine that the network is larger
when all method components and relations from~\cite{sanden97} are
drawn, and that the network is even larger when method
components and relations from other documents are present.

Now the importance of a unified framework comes into view: whenever
the professionals filling the library do not use the same framework, the
network cannot be constructed automatically. 

In addition, it can be seen why adding a 
new document is a non-trivial task: the
relation with all the existing method components must be identified.
Also, the correspondence between method components from the new document and
already existing ones must also be identified. See for example the
method component ``job model''. This relation was explicit 
in~\cite{sanden97} so it could be identified easily. But it could have been
that this document was already represented in the library. In that case the
the two networks would have been linked.

\SubSection{Using the library}
In section~\ref{sec:uselib} the various access mechanisms to the
architecture method library have been discussed. For answering exact
queries the method components are accessed directly using some kind of
query language. Query by example can be done using networks as discussed
in the previous section by going from node to node using the relations
and marking the interesting ones. Access using a decision tree 
resembles using a network which has pre-marked nodes; the professional
will not ``see'' the non-marked nodes. Again, he or she can mark
the interesting ones.

After browsing the architecture method library, the expert obtains a
list of selected method components and relations between them. These can
then be used in e.g. a project plan.

\Section{Future directions}
Eventually, once the library has been filled with a reasonable number
of method components, a core body of knowledge for
information systems engineering may be identified. 
Such an explicit and accepted body of knowledge would, for example:
\begin{itemize}
   \item Allow universities and training institutes to tune their
     curricula to a well defined body of knowledge accepted by both
     industry and academia.
   \item Allow for the identification of distinct roles in the
     information system engineering process and related forms of
     certification.
   \item Allow managers of information systems engineering projects to
     constitute project teams with professionals who share a common
     terminology and understanding of the profession.
   \item Allow client organizations to organize second opinion reviews
     among providers of information system engineering.
   \item Allow for re-use of experiences and materials between
     practitioners from different backgrounds.
\end{itemize}

Ideally, the library
should make use of a common and unified conceptual framework.
This will require several iterations of analysis of the library. 

Once the library becomes filled with some critical mass of assets, 
it may become part of a ``learning cycle''.  The actors in an
information systems engineering project who use the
knowledge in the library for their project activities will be able to
give feedback on the contents of the library based on their
experiences in a project.

Of course, these experiences should be added to the library,
resulting in an open library where
professionals submit method components or refinements of existing
ones to the library.  

\Section{Conclusion}
In this paper we have presented an architecture method library which
eventually may evolve to some kind of
information systems engineering body of knowledge. 
Discussed are the restrictions on the scope of the library, the contents
of the library, how the library can be used and how it can be filled, and
finally future directions. Also, an extended example has been provided.

This project is a collaboration between the Dutch Ordina group and the
University of Nijmegen. We are currently in the process of gathering
information systems engineering knowledge within the context of the Ordina
group. We appreciate contributions of other companies, research groups
and universities and invite them to join our
effort in gathering and structuring the architecture method library.

\bibliographystyle{alpha}
\bibliography{all}

\end{document}

%% file: pub.tex
{\sc Published as:}
\begin{quote}
  R.D.T. {Janssen}, H.A.~(Erik) {Proper}, H.~{Bosma}, D.~{Verhoef}, and S.J.B.A. {Hoppenbrouwers}. {Developing an Architecture Method Library}. Technical report, Ordina Institute, Gouda, The Netherlands, EU, January 2001.
\end{quote}